\documentclass{article}
\usepackage{ijcai18}

\usepackage{times}
\usepackage{xcolor}
\usepackage{soul}
\usepackage[small]{caption}
\usepackage{subfigure}
\usepackage{graphicx}
\usepackage{amsthm}
\usepackage{epsfig,algorithmic,algorithm}
\usepackage{graphicx}
\usepackage{subfigure}
\usepackage{multirow}

\usepackage{amsmath,bm}
\usepackage{amssymb}

\usepackage[english]{babel}

\usepackage{arydshln}

\usepackage[vmargin=1.2cm,hmargin=2.1cm,head=30pt,includeheadfoot]{geometry}
\usepackage{fancyhdr}
\title{A Brand-level Ranking System with the Customized Attention-GRU Model\thanks{Deng Cai is the corresponding author}}

\author{Yu Zhu$^{1 2}$, Junxiong Zhu$^3$, Jie Hou$^3$, Yongliang Li$^3$, Beidou Wang$^1$, Ziyu Guan$^4$, Deng Cai$^{1 2}$\\
$^1$State Key Lab of CAD\&CG, College of Computer Science, Zhejiang University, China\\
$^2$Alibaba-Zhejiang University Joint Institute of Frontier Technologies\\
$^3$Alibaba Group, Hangzhou, China\\
$^4$School of Information and Technology, Northwest University of China\\
\{zhuyu\_cad, dcai\}@zju.edu.cn, 
xike.zjx@taobao.com, 
muhuo.hj@alibaba-inc.com, 
\\anthonylee.liyl@tmall.com, 
beidouw@sfu.ca, 
ziyuguan@nwu.edu.cn
}
\begin{document}
\thispagestyle{fancy}
\chead{\LARGE \textbf{\emph{Published in IJCAI 2018}}} 
\rhead{}
\lhead{}
\maketitle

\begin{abstract}
  In e-commerce websites like Taobao, brand is playing a more important role in influencing users' decision of click/purchase, partly because users are now attaching more importance to the quality of products and brand is an indicator of quality. However, existing ranking systems are not specifically designed to satisfy this kind of demand. Some design tricks may partially alleviate this problem, but still cannot provide satisfactory results or may create additional interaction cost. In this paper, we design the first brand-level ranking system to address this problem. The key challenge of this system is how to sufficiently exploit users' rich behavior in e-commerce websites to rank the brands. In our solution, we firstly conduct the feature engineering specifically tailored for the personalized brand ranking problem and then rank the brands by an adapted Attention-GRU model containing three important modifications. Note that our proposed modifications can also apply to many other machine learning models on various tasks. We conduct a series of experiments to evaluate the effectiveness of our proposed ranking model and test the response to the brand-level ranking system from real users on a large-scale e-commerce platform, i.e. Taobao.
\end{abstract}

\section{Introduction}
\begin{figure}[htb!]
\begin{center}
\subfigure[Results of existing ranking systems]{\includegraphics[width=0.85\columnwidth]{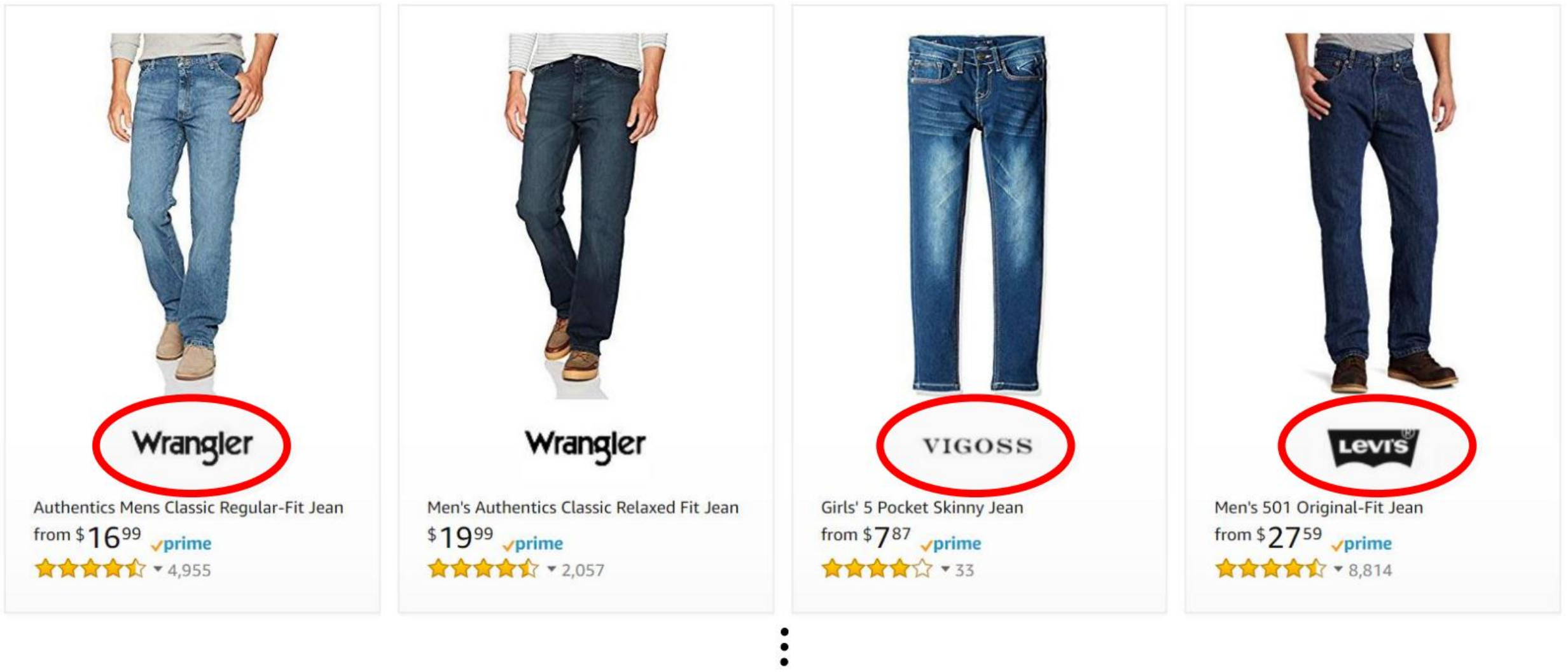}}
\subfigure[Results after clicking the checkbox Levi's]{\includegraphics[width=0.85\columnwidth]{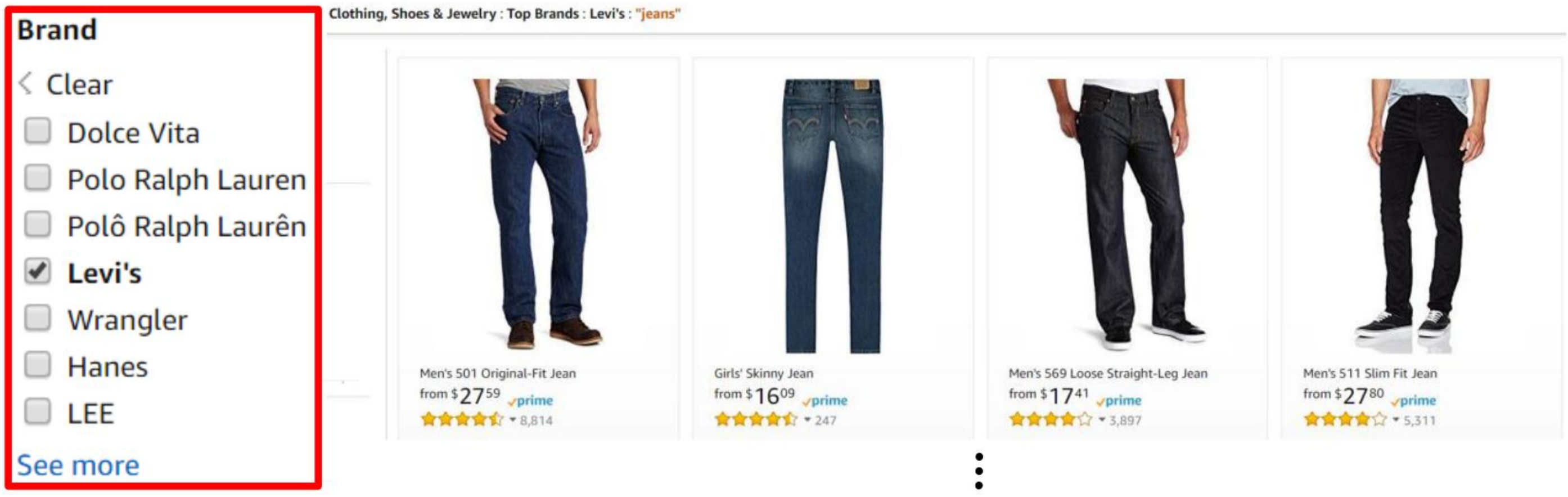}}
\subfigure[Results of our brand-level ranking system]{\includegraphics[width=0.78\columnwidth]{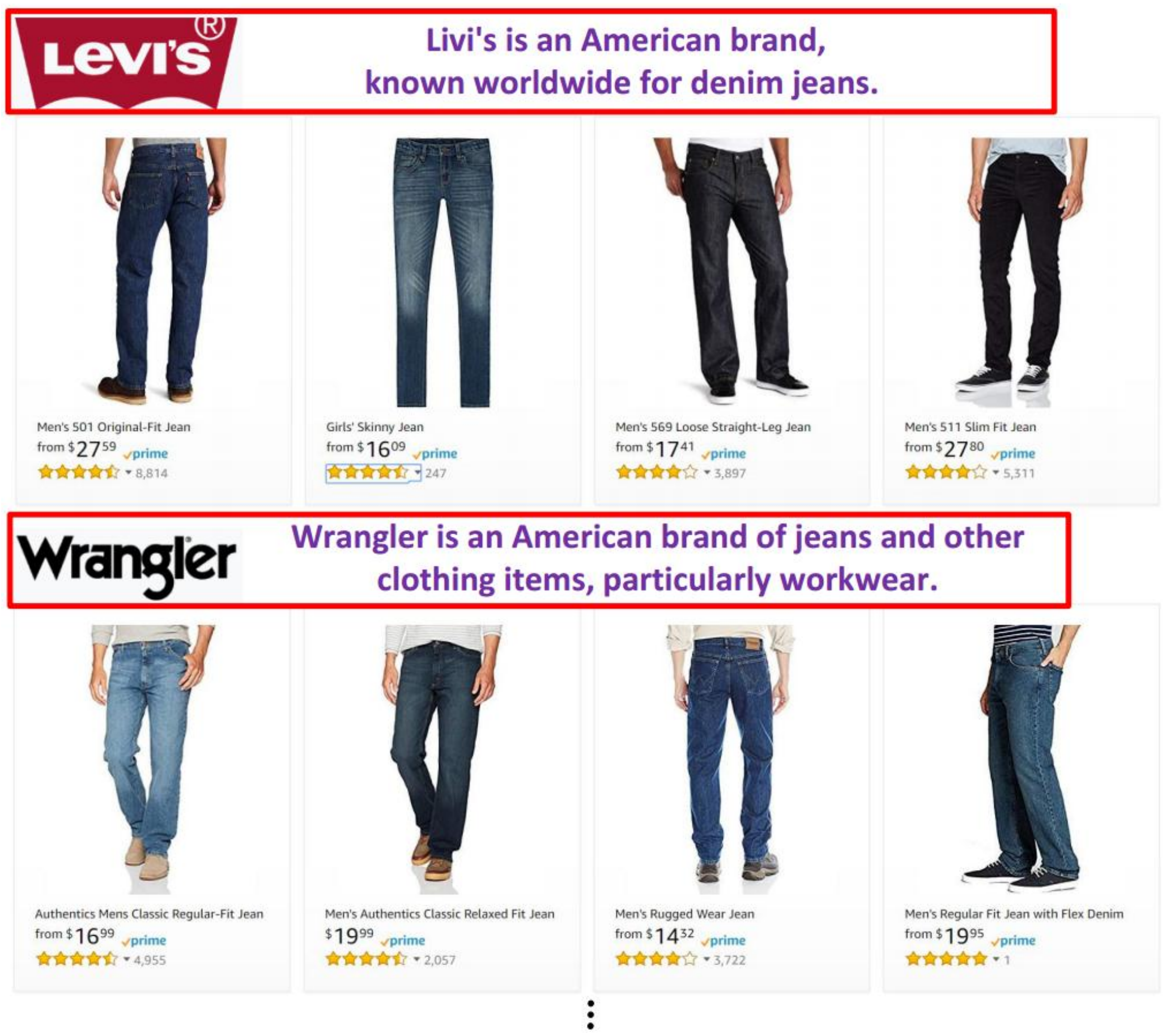}}
\end{center}
\vspace*{-15pt}
   \caption{Results of different ranking systems given the query “jeans”. Red lines highlight the important brand-related information.
}
\label{fig:motivation example}
\vspace*{-17pt}
\end{figure}
In e-commerce websites such as Taobao, the brand is having a growing impact on customers' decision of click/purchase \cite{zipser:2016modernization}. There are numerous reasons behind this phenomenon. On one hand, users may prefer items with high quality, and brand is a good indicator of product quality. On the other hand,  users may have preference bias to certain brands because of the brand image built from brand marketing and campaigns, e.g. basketball fans may prefer clothes under the brand endorsed by NBA stars.

However, as shown in Figure \ref{fig:motivation example}(a), the ranking results from most existing retrieval/recommender systems aggregate the items of different brands together. This cannot well satisfy the users with strong preference towards certain brands, since they have to waste time browsing through large numbers of products from other brands. There exist several technical tricks to help with this issue: (1) Set the brand as a feature for the underlying ranking model, and then items with preferred brands would rank in the top of the list. However, items of different brands are still mixed together. (2) Set checkboxes/buttons to filter brands. As shown in Figure \ref{fig:motivation example}(b), on the result page of query ``jeans", users can choose to only browse products under brand Levi's by clicking the checkbox filter for Levi's. However, only a few brands can be displayed here and there is no personalization. Moreover, users have to click the checkboxes multiple times if they want to browse several brands, which increases the interaction cost and creates negative user experience. 

Due to the drawbacks of existing systems mentioned above, we design a new brand-level ranking system, where items of the same brand are grouped together and brand groups are ranked based on users' brand preference. A demo of our system is demonstrated in Figure \ref{fig:motivation example}(c). Jeans are first grouped by brands, e.g. Levi's and Wrangler. The system will then rank Levi's at a higher position, if it learns the user prefers Levi's to Wrangler.

The core of the brand-level ranking system is to solve the personalized brand ranking problem. Various important information in e-commerce websites should be exploited to rank brands. Firstly, features such as brand pricing contribute to brand profiling. Second, the sequential information in user action (\emph{click} or \emph{purchase}) sequences and time intervals between actions are valuable in modeling users' interests on brands \cite{zhu:2017next}. In addition, different types of actions (\emph{click} and \emph{purchase}) reflect users' different preferences on items (and in turn the corresponding brands), e.g. generally a user purchasing an item indicates he is more interested in the item than if he clicks it. The key challenge is how to well exploit these characteristics of user action sequences for brand ranking. We formulate it as a point-wise ranking problem. Specifically, a classifier is trained based on the various information in e-commerce websites. Then given a user, his probability of preferring each brand is predicted by the classifier and brands are ranked based on their probabilities. Items within the same brand group can be ranked by traditional ranking methods.

Two sub-tasks need to be accomplished for the brand-level ranking  system. One is the feature engineering for the brand ranking problem. We propose a series of brand features that are important for brand ranking (details in section 4.1). The other one is the design of the ranking model. We propose an adapted Attention-GRU \cite{chorowski:2015attention} model to predict users' probabilities of preferring different brands. RNN (Recurrent Neural Network) methods have achieved state-of-the-art performance in capturing the sequential information in user action sequences \cite{zhu:2017next}. GRU (Gated Recurrent Unit) \cite{cho:2014learning} is among the best RNN architectures and the attention mechanism \cite{chorowski:2015attention} helps to distinguish the influence of different previous actions on the current prediction. Therefore, Attention-GRU is chosen as the base model. We propose three major modifications to Attention-GRU for our task. (1) Combine the heuristically designed brand features and the model-learned brand embedding to better represent the brand. A brand embedding refers to a vector representation for the brand, which is learned from training data by our model \cite{zhu:2016heterogeneous}. (2) Consider different types of actions. (3) Integrate the time-gate \cite{zhu:2017next} to model time intervals between actions, which can better capture users' short-term and long-term interests. The effectiveness of our adapted model and the brand-level ranking system is evaluated in our offline and online experiments on a large-scale e-commerce platform.
This paper's contributions are outlined as follows.
\begin{itemize}
\item We propose the first brand-level ranking system, which provides explicit personalized brand-level ranking and better helps users to make click/purchase decisions based on their brand preference.
\item We perform feature engineering tailored for the brand ranking task and propose an adapted Attention-GRU model as our ranking model. Specifically, we contribute three modifications which effectively improve the model performance.
\item We conduct extensive offline and online experiments on a large-scale e-commerce platform. The results and feedbacks from real users prove the effectiveness of our adapted model and the brand-level ranking system.

\end{itemize}

\section{Related Work}

\subsection{RNN, GRU and Attention-GRU}
\emph{RNN} \cite{elman:1990finding} has been proved to perform excellently when modeling sequential data. It is formally defined as:
\begin{align}
&s_m = f(Wx_m + Us_{m-1})\label{eq:RNN1},\\
&o_m = softmax(Vs_m)\label{eq:RNN2},
\end{align}
where $s_m$, $x_m$ and $o_m$ are the hidden state, input and output at the $m$-th step. $f$ is a non-linear activation function. $W$, $U$ and $V$ are the corresponding weights learned from training.

\emph{GRU} \cite{cho:2014learning} is an important ingredient of RNN architectures, which can avoid the problem of gradient vanishing. It replaces Eq. (\ref{eq:RNN1}) with:
\begin{align}
&z_m = \sigma(W_zx_m + U_zs_{m-1})\label{eq:GRU1},\\
&r_m = \sigma(W_rx_m + U_rs_{m-1})\label{eq:GRU2},\\
&s_m = z_m\odot tanh(W_hx_m + U_h(r_m\odot s_{m-1}))\notag\\
&\mbox{~~~~~~~~~}+(1 - z_m)\odot s_{m-1}\label{eq:GRU3},
\end{align}
where $z_m$ and $r_m$ are the update and reset gates. $\sigma$ and $tanh$ are sigmoidal and tanh nonlinearities. $W_z$, $U_z$, $W_r$, $U_r$, $W_h$ and $U_h$ are the weights. $\odot$ denotes the element-wise product.

\emph{Attention-GRU} \cite{chorowski:2015attention} refers to GRU with the attention mechanism. RNN methods with the attention mechanism have been successfully applied to handwriting synthesis \cite{graves:2013generating}, machine comprehension \cite{pan:2017memen} etc. It typically focuses on the task that generates an output sequence $y = (y_1,\cdots, y_T)$ from an input sequence $x = (x_1,\cdots, x_L)$. $x$ is often processed by an \emph{encoder} to output a sequential representation $h=(h_1,\cdots, h_L)$. At the $m$-th step $y_m$ is generated by:
\begin{align}
&\alpha_m = Attend(s_{m-1}, h)\label{eq:Attention-GRU1},\\
&g_m = \sum_{j=1}^L\alpha_{m,j}h_j\label{eq:Attention-GRU2},\\
&s_m = Recurrency(y_{m-1},s_{m-1},g_m)\label{eq:Attention-GRU3},\\
&y_m\sim Generate(y_{m-1},s_m,g_m)\label{eq:Attention-GRU4},
\end{align}
where $Attend$ and $Generate$ are functions. $\alpha_m$ is a vector whose entry $\alpha_{m,j}$ indicates the \emph{attention weight} of the $j$-th input. $g_m$ is called a \emph{glimpse} \cite{mnih:2014recurrent}. $Recurrency$ represents the recurrent activation. In Attention-GRU, the recurrent activation is GRU. 
\subsection{RNN for Behavior Modeling}
\cite{hidasi:2015session,tan:2016improved} focus on RNN solutions in a session-based setting. \cite{yu:2016dynamic} designs an RNN method for the next-basket recommendation. The setting in our task is different from these settings. The Time-LSTM model, proposed in \cite{zhu:2017next}, equips LSTM \cite{hochreiter:1997long} with time gates to better capture users' short-term and long-term interests. We propose that the time gate is a basic component and can be integrated into other RNN architectures (e.g. Attention-GRU in our task).
\begin{figure}[tb]
\begin{center}
\includegraphics[scale=0.28]{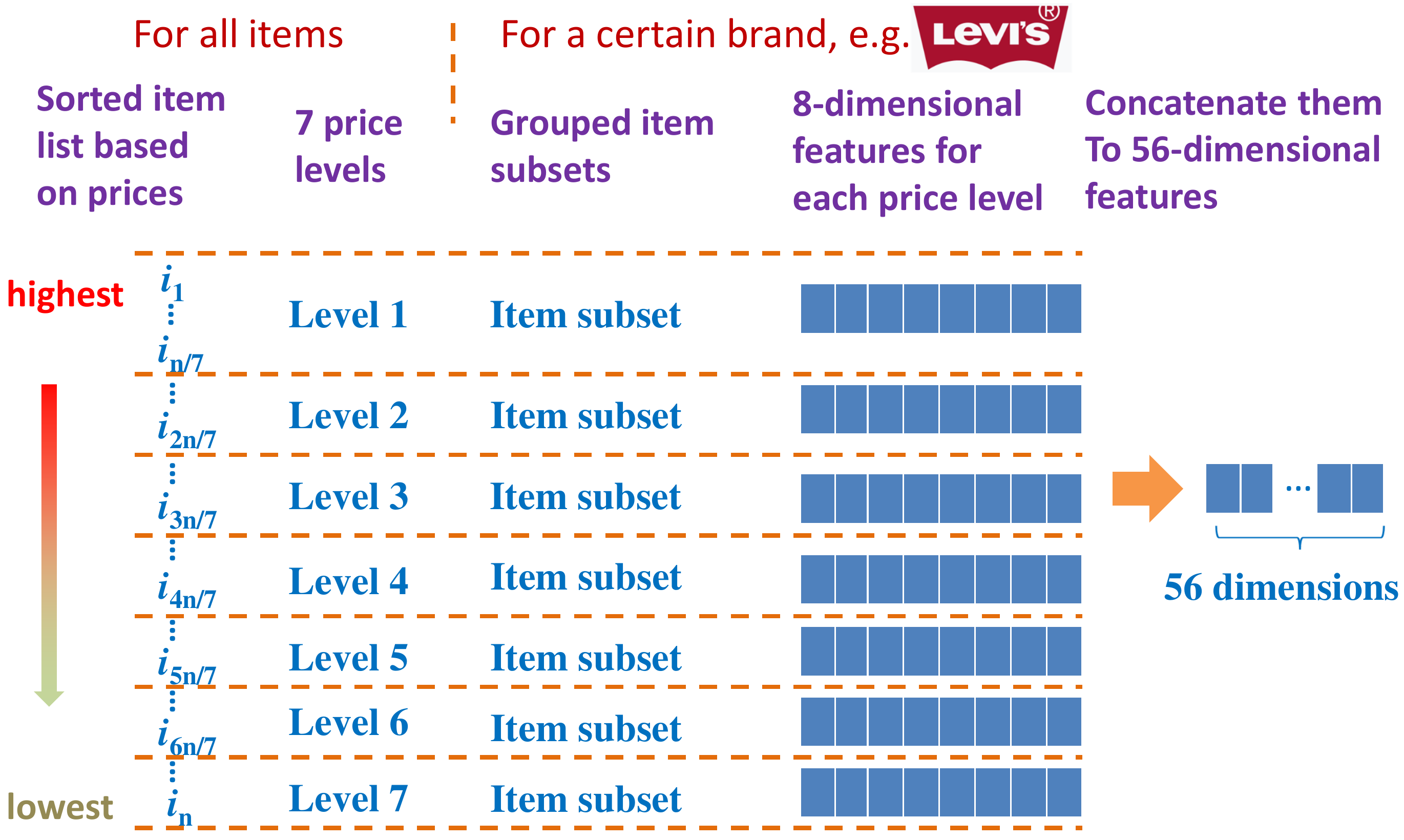}
\end{center}
\vspace*{-10pt}
   \caption{The general process to obtain the brand features.
}
\label{fig:feature engineer}
\vspace*{-15pt}
\end{figure}

\section{Task Definition and Models' adaptations}
\subsection{Task Definition}
Let $\mathbf{U} = \{U_1, U_2, \cdots, U_M\}$ be a set of $M$ users and $\mathbf{B} = \{B_1, B_2, \cdots, B_N\}$ be a set of $N$ brands. For each user $u\in \mathbf{U}$, his behavior history $H^u$ is defined by $H^u=[(b_1^u,h_1^u,t_1^u), (b_2^u,h_2^u,t_2^u), \cdots, (b_{n_u}^u,h_{n_u}^u,t_{n_u}^u)]$, where $(b_m^u,h_m^u,t_m^u)$ indicates that $u$ generates his $m$-th action $h_m^u$ (\emph{click} or \emph{purchase}) on brand $b_m^u\in \mathbf{B}$ at time $t_m^u$. Our task is to predict the probability $p(B_{q_u})$ of $u$ generating any type of action on brand $B_{q_u}$ at a certain time $T_u$ ($T_u>t_{n_u}^u$).
\subsection{Adaptations of Traditional RNN Models}
We first transform $H^u$ to $\tilde{H^u} = [(b_1^u,h_1^u,t_2^u-t_1^u), (b_2^u,h_2^u,t_3^u$ $-t_2^u), \cdots, (b_{n_u}^u,h_{n_u}^u,T_u-t_{n_u}^u)]$. The mathematical representations of $b_m^u$, $h_m^u$ and $t_{m+1}^u-t_m^u$ are denoted as $R(b_m^u)$, $R(h_m^u)$ and $R(t_{m+1}^u-t_m^u)$, respectively. $R(b_m^u)$ is defined with two alternative ways: (1) a brand feature vector or (2) a one-hot vector. $R(h_m^u)$ is defined as a one-hot vector. Since there are two types of actions, i.e. \emph{click} and \emph{purchase}, in our task, thus they are represented as [0, 1] and [1, 0], respectively. $R(t_{m+1}^u-t_m^u)$ is set to be a scalar $t_{m+1}^u-t_m^u$ (in seconds). The label is $1$ if $u$ actually clicks/purchases $B_{q_u}$ at time $T_u$ and $0$ otherwise.

For Time-LSTM \cite{zhu:2017next}, the input at each step is the concatenation $[R(b_m^u), R(h_m^u)]$ and $R(t_{m+1}^u-t_m^u)$. For the other RNN models, the input (i.e. $x_m$ in section 2.1) is the concatenation $[R(b_m^u), R(h_m^u), R(t_{m+1}^u-t_m^u)]$. For RNN with no attention mechanism, the output (i.e. $o_m$ in Eq. (\ref{eq:RNN2})) is a probability distribution over all brands, from which we can obtain $p(B_{q_u})$. For Attention-GRU, $T$ and $y_0$ (i.e. $y_{m-1}$ in Eq. (\ref{eq:Attention-GRU3}) when $m=1$) in section 2.1 are set to be $1$ and $B_{q_u}$, respectively. We replace Eq. (\ref{eq:Attention-GRU4}) with:
\begin{align}
&\tilde{o}_m = softmax(\tilde{V}s_m)\label{eq:Output_GRU}, 
\end{align}
where $\tilde{V}$ is the weight and $\tilde{o}_m$ is a probability distribution over labels $1$ and $0$. $p(B_{q_u})$ is equal to the probability of label $1$. These models are trained by AdaGrad \cite{duchi:2011adaptive} with the log loss calculated by $p(B_{q_u})$ and the label.

\begin{table}[tb] 
\scriptsize
\centering
\begin{tabular}{|c|c|c|}
\hline
 CTR (Click-Through Rate)&CVR (Conversion Rate)\\ \hline
 GMV (Gross Merchandise Volume)&ATIP (Average Transacted Item Price)\\ \hline
 Search Times&Click Times\\ \hline
 Add-To-Cart Times&Transaction Times\\ \hline
\end{tabular}\vspace*{-5pt}\caption{8 Most Important E-commerce Metrics}
\vspace*{-12pt}
\label{table:8 metrics}
\end{table} 
\section{Proposed System}

\subsection{Feature Engineering}
According to the market research conducted by the authors' company, the price range feature demonstrates great importance in user and brand profiling. Therefore, the brand features are carefully engineered by slicing into 7-level price ranges, with the details as follows (also shown in Figure \ref{fig:feature engineer}).


(1) All items belonging to a certain category (e.g. ``clothing") are sorted based on their prices. The sorted item list is denoted as $\{i_1,i_2,\cdots,i_n\}$. The price for $i_t$ is denoted as $p_t$. We define $7$ \textit{price levels} as 7 price ranges, with Level 1 = $(0, p_\frac{n}{7}]$, Level 2 = $(p_\frac{n}{7},p_\frac{2n}{7}]$,$\cdots$, Level 7 = $(p_\frac{6n}{7},p_n]$.

(2) Given a certain brand, items belonging to this brand are grouped based on the price levels defined above. Specifically, an item is grouped to a price level if this item's price locates in the corresponding price range.

(3) Given a certain brand, 8 most important e-commerce metrics are aggregated within each price level, generating 8-dimensional features for each price level. Features from all 7 price levels are then concatenated into the final 56-dimensional brand features. Specifically, the 8 most important brand-related e-commerce metrics used in this paper is summarized in \emph{Table \ref{table:8 metrics}}\footnote{Due to page limit, please refer to the corresponding Wikipedia pages for definitions of CTR, CVR and GMV. ATIP = $ \frac{\mbox{GMV}}{\mbox{Number of Transacted Items}}$.}.

\subsection{The Design of the Ranking Model}
An adapted Attention-GRU model is proposed to predict users’ probabilities of preferring different brands.

\subsubsection{(1) The Choice of Attention-GRU}
\begin{figure}[tb]
\begin{center}
\includegraphics[scale=0.27]{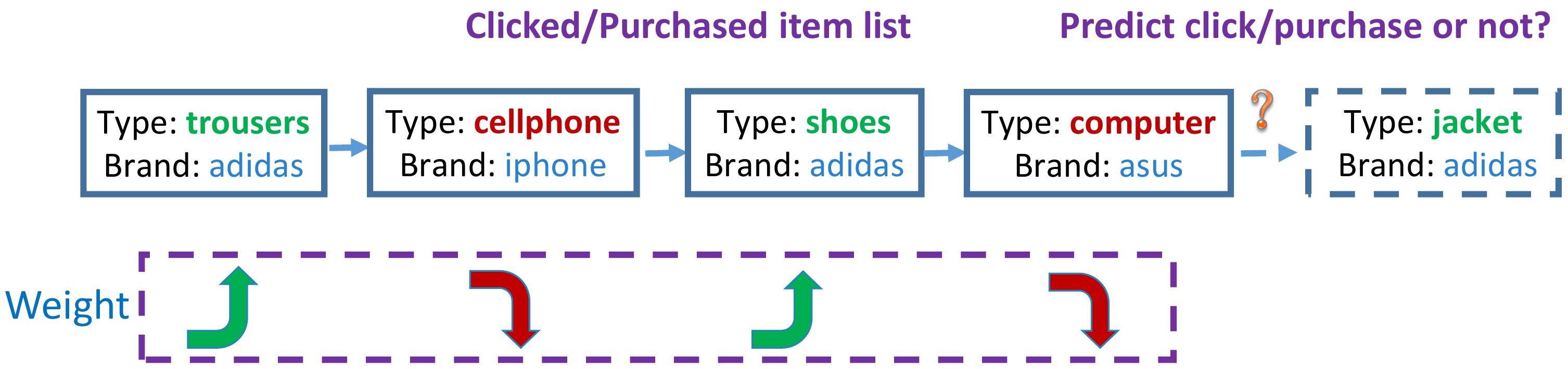}
\end{center}
\vspace*{-10pt}
   \caption{An example to show the motivation to apply the attention mechanism in our task.
}
\label{fig:why Attention-GRU}
\vspace*{-12pt}
\end{figure}

RNN solutions have achieved state-of-the-art performance when modeling users' sequential actions \cite{hidasi:2015session,zhu:2017next}. LSTM and GRU are two important ingredients of RNN architectures, both of which are able to avoid the problem of gradient vanishing. Compared to LSTM, GRU can yield comparable performance but has fewer parameters, thus it could train a bit faster and need less data to generalize \cite{chung:2014empirical}. Figure \ref{fig:why Attention-GRU} is an example to show the motivation to apply the attention mechanism in our task. When we predict whether the user would click/purchase the jacket with the brand Adidas, we want to increase the influence of the first and third items on the prediction since compared to the cellphone and computer, the trousers and shoes are much more related to the jacket. In the attention mechanism, this will be reflected with larger weights on the first and third inputs.

\subsubsection{(2) Modifications to Attention-GRU}
To achieve better performance, we propose three modifications to Attention-GRU (also explained in Figure \ref{fig:ThreeModification}):
\begin{figure*}[tb]
\begin{center}
\includegraphics[scale=0.38]{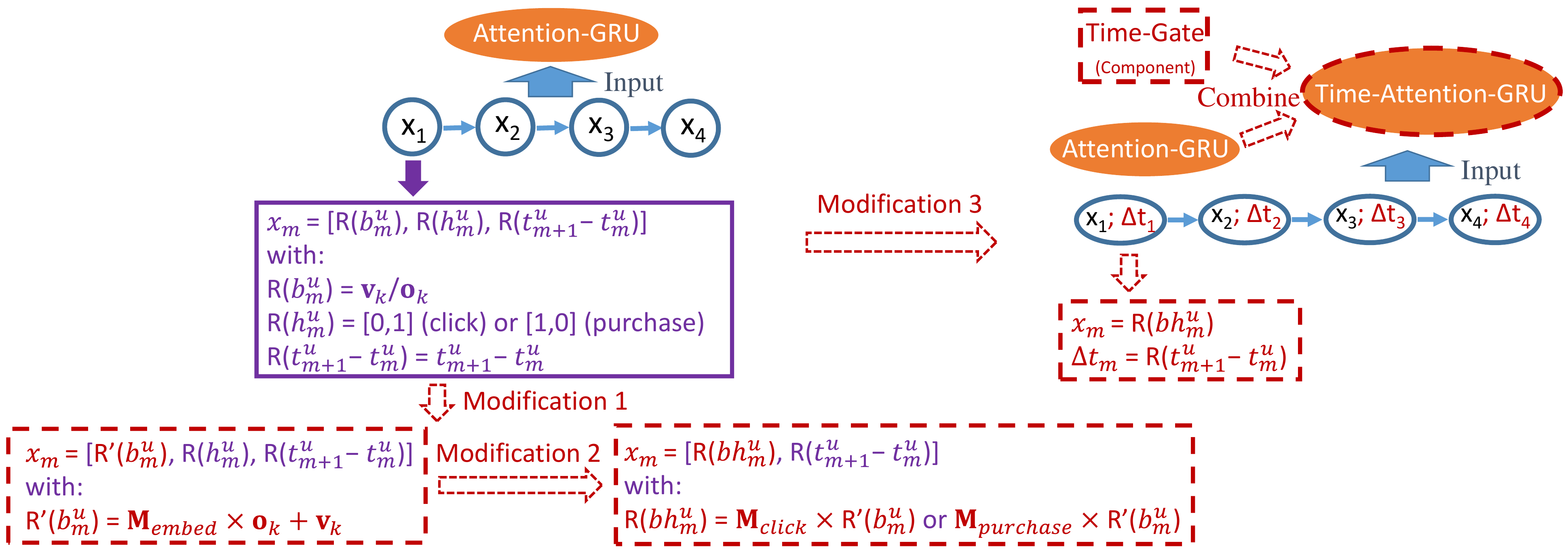}
\end{center}
\vspace*{-10pt}
   \caption{Our proposed modifications to the Attention-GRU model, represented by dotted lines and red texts.
}
\label{fig:ThreeModification}
\vspace*{-15pt}
\end{figure*}

\noindent \textbf{Modification 1: Combining the brand features and brand embedding to better represent the brand}\\
In a traditional Attention-GRU model, the brand can be represented by either a vector consisting of brand features defined in section 4.1, or a one-hot vector based on brand ID. However, the brand features are heuristically designed based on expert experience and may lack some important information useful for the brand ranking task. For example, two brands may have similar brand features but are of different interest for the same user. Using one-hot vectors can explicitly distinguish different brands but lacks the brand content information. Here we propose to combine these two representations. Specifically, following the denotations in section 3.1, assume $b_m^u$ equals to $B_k$ and the vector consisting of brand features for $B_k$ is denoted as $\mathbf{v}_k\in R^{56\times 1}$. Its one-hot vector is defined as $\mathbf{o}_k\in \{0,1\}^{N\times 1}$, with the $k$-th entry equal to  $1$ and the other entries equal to $0$. Matrix $\mathbf{M}_{embed} \in R^{56\times N}$ is defined to contain the embeddings of all brands. Then the combined  mathematical representation considering both brand features and the brand embedding is defined as   $R^\prime(b_m^u)$, with 
\begin{equation}\label{eq:combine}
\begin{aligned}
&R^\prime(b_m^u) = \mathbf{M}_{embed}\times \mathbf{o}_k + \mathbf{v}_k.
\end{aligned}
\end{equation}
$\mathbf{M}_{embed}$ is learned from the training data. $\mathbf{o}_k$ is used to look up the embedding of $B_k$ from $\mathbf{M}_{embed}$.

We interpret Eq. (\ref{eq:combine}) from three perspectives. \emph{(1) Heuristic initialization + Fine tuning}. While $\mathbf{v}_k$ is heuristically designed, by learning from the training data, we fine tune the brand's representation to be $\mathbf{M}_{embed}\times \mathbf{o}_k + \mathbf{v}_k$, which would much better fit our task. \emph{(2) Prior information + Posterior modification}. This perspective is similar to the first perspective, but in a Bayesian context \cite{berger:2013statistical}. Specifically, $\mathbf{v}_k$ is obtained based on our ``prior'' understanding of brands, then the training data enhances our understanding of them and allows us to give ``posterior'' estimations of their representations. \emph{(3) Content information + Collaborative information}. $\mathbf{v}_k$ contains brands' content information, while $\mathbf{M}_{embed}\times \mathbf{o}_k$ is learned from users' click/purchase data in our task and thus captures the collaborative information. For some brands $B_k$ that rarely appear in the training data, $\mathbf{v}_k$ can provide extra information to help with this ``cold-start'' situation \cite{wang:2016view,wang:2016million}.

\noindent \textbf{Modification 2: Considering different types of actions}\\
When modeling user behaviors by RNN models, previous works such as \cite{hidasi:2015session,zhu:2017next} usually consider single type of action. However, users can perform multiple types of actions on items, e.g. \emph{click} and \emph{purchase}. One way to handle different types of actions is to model each of them as a one-hot vector $R(h_m^u)$ (defined in section 3.2) and concatenate it to our combined brand representation $R^\prime(b_m^u)$. However, this method allows no feature interaction between $R^\prime(b_m^u)$ and $R(h_m^u)$ (or implicit interaction by the non-linear activation function), and thus would not well capture the interactive relation between the brand and the type of action performed on the brand \cite{rendle:2012factorization,chen:2017user}.

Instead, we propose to define a matrix for each type of action (the total number of action types is usually not large). In our task, we define $\mathbf{M}_{click}\in R^{56\times 56}$ and $\mathbf{M}_{purchase}\in R^{56\times 56}$ for \emph{click} and \emph{purchase}, respectively. Then the representation considering both of the brand and the type of action is defined as:
\begin{equation}\label{eq:Brand considering actions}
   R(bh_m^u)=
   \begin{cases}
   \mathbf{M}_{click}\times R^\prime(b_m^u), &if\  b_m^u\   is\  \emph{clicked},\\
   \mathbf{M}_{purchase}\times R^\prime(b_m^u), &if\ b_m^u\ is\   \emph{purchased}.
   \end{cases}
\end{equation}
In this way, we explicitly model the interaction between the brand and the type of action by matrix multiplication.

\noindent \textbf{Modification 3: Integrating the time-gate to model time intervals between actions}\\
The time gate in Time-LSTM \cite{zhu:2017next} is effective to capture users' short-term and long-term interests. We propose that the time gate is a basic component and adapt it to our model as follows. A time gate $T_m$ is defined as:
\begin{equation}\label{eq:Time gate}
\begin{aligned}
&T_m = \sigma (W_tx_m + \sigma(Q_{t}\triangle t_m)),
\end{aligned}
\end{equation}
where
\begin{equation}\label{eq:Input of Time-Attention-GRU}
\begin{aligned}
&x_m = R(bh_m^u),\\
&\triangle t_m = R(t_{m+1}^u-t_m^u) = t_{m+1}^u-t_m^u.
\end{aligned}
\end{equation}
$W_t$ and $Q_t$ are the weights of $x_m$ and $\triangle t_m$ (time intervals), respectively. $\triangle t_m$ may be very large, thus we impose a sigmoidal function $\sigma$ on $Q_{t}\triangle t_m$. Then we modify Eq. (\ref{eq:GRU3}) to:
\begin{equation}\label{eq:GRU_Timegate}
\begin{aligned}
&s_m = z_m\odot T_m\odot tanh(W_hx_m + U_h(r_m\odot s_{m-1}))\\
&\mbox{~~~~~~~~~}+(1 - z_m)\odot s_{m-1}.
\end{aligned}
\end{equation}
When modeling user behaviors by our model, $x_m$ in Eq. (\ref{eq:GRU_Timegate}) represents the user's most recent action, thus we can exploit $x_m$ to learn his/her current short-term interest. $s_{m-1}$ models this user's previous actions, thus $s_{m-1}$ reflects his/her long-term interest. $T_m$ is helpful in two ways, i.e. (1) $tanh(W_hx_m + U_h(r_m\odot s_{m-1}))$ is filtered by not only $z_m$, but also $T_m$. So $T_m$ can control the influence of $x_m$ on current prediction. (2) $\triangle t_m$ is firstly stored in $T_m$, then transferred to $s_m$, and would be transferred to $s_{m+1}, s_{m+2}\cdots$. Thus $T_m$ helps to better model users' long-term interest ($s_m, s_{m+1}\cdots$) by incorporating $\triangle t_m$ for later predictions. We denote Attention-GRU with the time gate as Time-Attention-GRU.

It is notable that Modification $1$ is rather general and can apply to many machine learning models where the designed features and one-hot vectors exist. Modification $2$ can be generalized to handle tasks involving various interaction types with the input. In Modification $3$, we propose that the time gate can be integrated into not only LSTM but also other RNN models, just as how we integrate it into Attention-GRU.

\subsubsection{(3) Loss and Training}

Similar to Attention-GRU, $p(B_{q_u})$ in our model is obtained from $\tilde{o}_m$ in Eq. (\ref{eq:Output_GRU}). We define the loss for user $u$ as:
\begin{equation}\label{eq:loss}
\begin{aligned}
&loss_u
   =\begin{cases}
   -log(p(B_{q_u})), & \mbox{if $label_u=1$},\\
   -w\times log(1 - p(B_{q_u})), & \mbox{otherwise},
   \end{cases}
\end{aligned}
\end{equation}
where $label_u$ is $1$ if $u$ actually generates an action on brand $B_{q_u}$ at time $T_u$. We multiply the loss by $w$ ($<1$) for negative instances because some unlabeled positive instances may be mixed in the training data \cite{pan:2008one}. Our model is optimized by AdaGrad \cite{duchi:2011adaptive}.


\section{Offline Experiments}
\subsection{Dataset}
A large-scale dataset is collected from Taobao. Specifically, we extract tuples $<$$user\_id$, $brand\_id$, $action\_type$, $timestamp$$>$, with each tuple representing that user $user\_id$ has an action $action\_type$ on brand $brand\_id$ at time $timestamp$. Users and brands with few interactions are filtered. The final data set in our offline experiments consists of $M = 3,591,372$ users, $N = 90,529$ brands and $82,960,693$ actions.

For each user $u$, his/her action sequence is cut into short ones with the length equal to $11$. The first $10$ actions form $H^u$. The brand and timestamp in the last action are assigned to $B_{q_u}$ and $T_u$ respectively, serving as the positive instances ($label_u=1$) in training. The negative instances ($label_u=0$) are generated by replacing $B_{q_u}$ with another random brand.
\subsection{Compared Models and Evaluations}
Our proposed model, Attention-GRU with three modifications, is denoted as Attention-GRU-3M, and is compared with the following baselines.\\
\noindent \textbf{\emph{GRU}:} GRU \cite{cho:2014learning} is among the best RNN architectures. Thus we choose GRU as a representative of the original RNN models.\\
\noindent \textbf{\emph{Attention-GRU}:} Similarly, Attention-GRU \cite{chorowski:2015attention} is selected to represent RNN models with the attention mechanism.\\
\noindent \textbf{\emph{Time-LSTM}:} Time-LSTM \cite{zhu:2017next} has achieved state-of-the-art performance for sequential behavior modeling. Thus it would be a competitive baseline in our task.\\
\noindent \textbf{\emph{Session-RNN}:} Session-RNN \cite{hidasi:2015session} exploits RNN to capture users' short-term interest based on sequential actions within a session. We use the publicly available python implementation\footnote{https://github.com/hidasib/GRU4Rec} of Session-RNN, with the session identified by \emph{timeout} \cite{huang:2004dynamic}.\\
\noindent \textbf{\emph{libFM}:} Our task can also be treated as a brand recommendation problem. Therefore, many hybrid recommendation methods that simultaneously capture the content and collaborative information can be applied. We use the popular libFM model \cite{rendle:2012factorization} as a representative of these methods.

GRU, Attention-GRU, Time-LSTM and Session-RNN are adapted to our task as described in section 3.2. For libFM, we extract tuples $<$$user\_id$, $brand\_id$, $label$$>$, where $label$ is $1$ if user $user\_id$ has an action on brand $brand\_id$ and $0$ otherwise. Then the one-hot vector for $user\_id$, the one-hot vector for $brand\_id$ and the corresponding brand feature vector are concatenated to be $\mathbf{x}$ and $label$ is $\mathbf{y}$ in libFM.
The number of units is empirically set to 256 for RNN models. The other hyperparameters in all models are tuned via cross-validation or set as in the original paper. Our code is public \footnote{https://github.com/zyody/Attention-GRU-3M}.

AUC and F1 score \cite{kim:2013nonparametric} are used to evaluate the prediction performance of different models.

\subsection{Results and Discussions}
\subsubsection{Model Comparison}
As shown in Table \ref{table:Method Comparison}, our proposed model significantly outperforms all baselines and we attribute it to the adoption of the attention mechanism and three proposed modifications. In comparison, baselines GRU and Session-RNN, exploiting none of these merits, perform the worst. Compared with GRU, Attention-GRU adopts additional attention mechanism and Time-LSTM adopts Modification 3 proposed in section 4.2. Thus, they both outperform GRU and Session-RNN. As a hybrid recommendation approach, libFM captures content and collaborative filtering features at the same time, but fails to capture the sequential information of user behaviors, while our model captures all these information. This explains why libFM performs worse than our model.

\begin{table}[t]
\scriptsize

\centering
\begin{tabular}{|c|c|c|c|c|c|}
\hline
 \multicolumn{2}{|c|}{}&\multicolumn{2}{|c|}{50\% training data}&\multicolumn{2}{|c|}{100\% training data}\\ \hline
 \multicolumn{2}{|c|}{}&AUC&F1&AUC&F1\\ \hline
\multicolumn{2}{|c|}{GRU}&0.5831&0.5573&0.5952&0.5752\\ \hline
\multicolumn{2}{|c|}{Attention-GRU}&0.5918&0.5698&0.6059&0.5843\\ \hline
\multicolumn{2}{|c|}{Time-LSTM}&0.5926&0.5704&0.6062&0.5849\\ \hline
\multicolumn{2}{|c|}{Session-RNN}&0.5839&0.5561&0.5967&0.5733\\ \hline
\multicolumn{2}{|c|}{libFM}&0.5891&0.5663&0.6009&0.5802\\ \hline
\multicolumn{2}{|c|}{Attention-GRU-3M}&\textbf{0.6051}$^*$&\textbf{0.5801}$^*$&\textbf{0.6283}$^*$&\textbf{0.6050}$^*$\\ \hline
\end{tabular}\vspace*{-5pt}\caption{Model Comparison (bold typeset indicates the
best performance. * indicates statistical significance at $p < 0.01$ compared to the second best.)}
\vspace*{-7pt}
\label{table:Method Comparison}
\end{table}

\begin{figure}[tb]
\begin{center}
\includegraphics[scale=0.3]{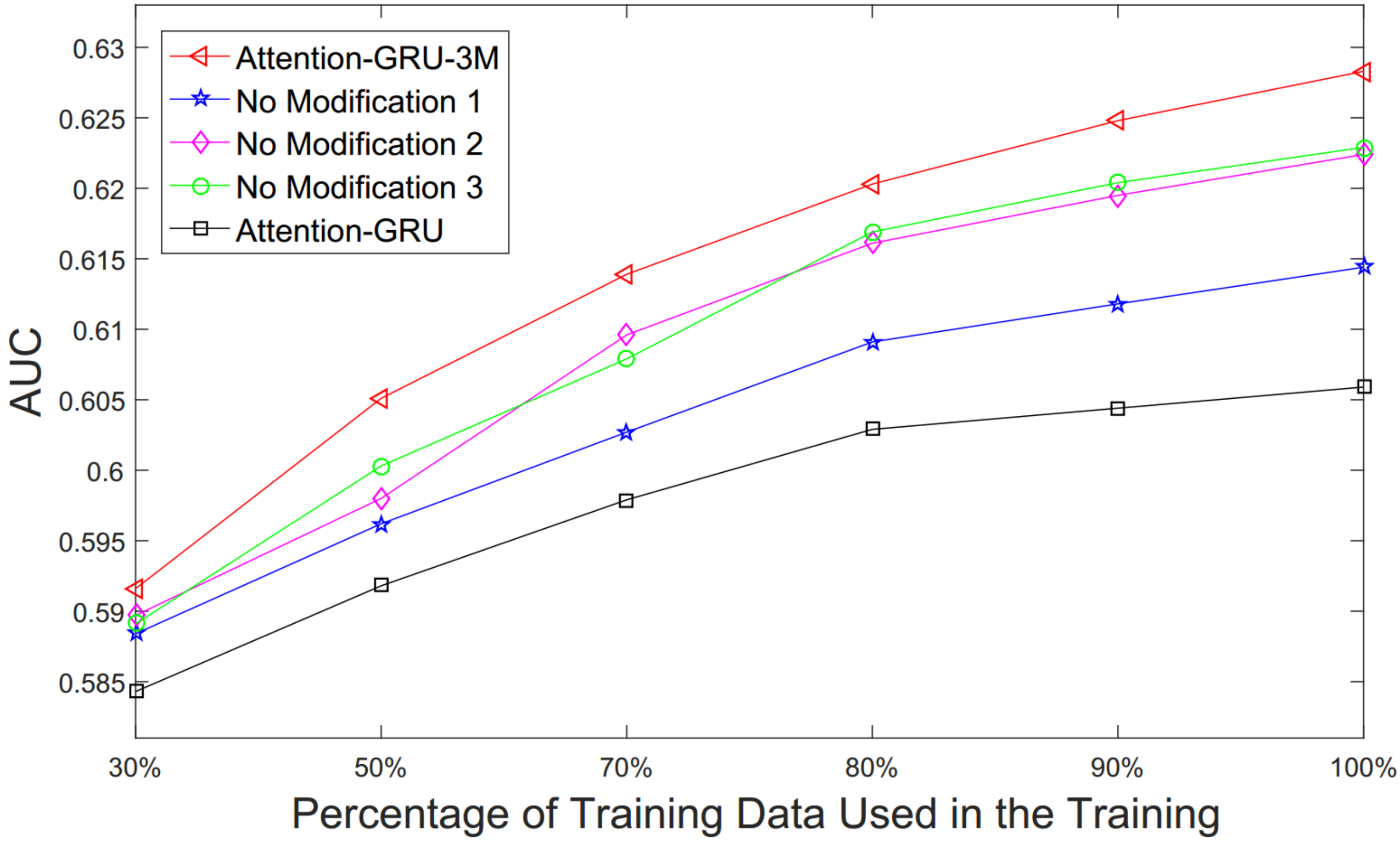}
\end{center}
\vspace*{-15pt}
   \caption{Performance measured by AUC when we remove three modifications one at a time and vary the training data size. No Modification $i$ refers to our model without Modification $i$.}
\label{fig:ThreeModifications}
\vspace*{-12pt}
\end{figure}

\subsubsection{The Effects of Three Modifications}
As described in section 4.2, three modifications are proposed to enhance Attention-GRU. We now remove them one at a time to evaluate how each modification affects the prediction performance. We also change the size of data used in training to evaluate its effect on the performance. As shown in Figure \ref{fig:ThreeModifications} (The results of AUC and F1 are similar. Due to page limit, we only show the results of AUC), AUC declines when we remove each type of modification and declines the most when removing Modification 1, which indicates that all the proposed modifications contribute to the performance improvement and Modification 1 is more effective. As the size of training data increases, the influence of Modification 1 becomes larger (i.e. the gap between ``Attention-GRU-3M" and ``No Modification 1" becomes larger). One explanation may be that, compared to the other two modifications, Modification 1 brings much more parameters (i.e. entries of $\mathbf{M}_{embed}$), thus it needs more training data to learn their optimal values. This also explains that in Table \ref{table:Method Comparison} and Figure \ref{fig:ThreeModifications}, our model has a larger performance improvement than the other models as the data size increases.

\section{Online Experiments}
\begin{figure}[tb]
\begin{center}
\includegraphics[scale=0.25]{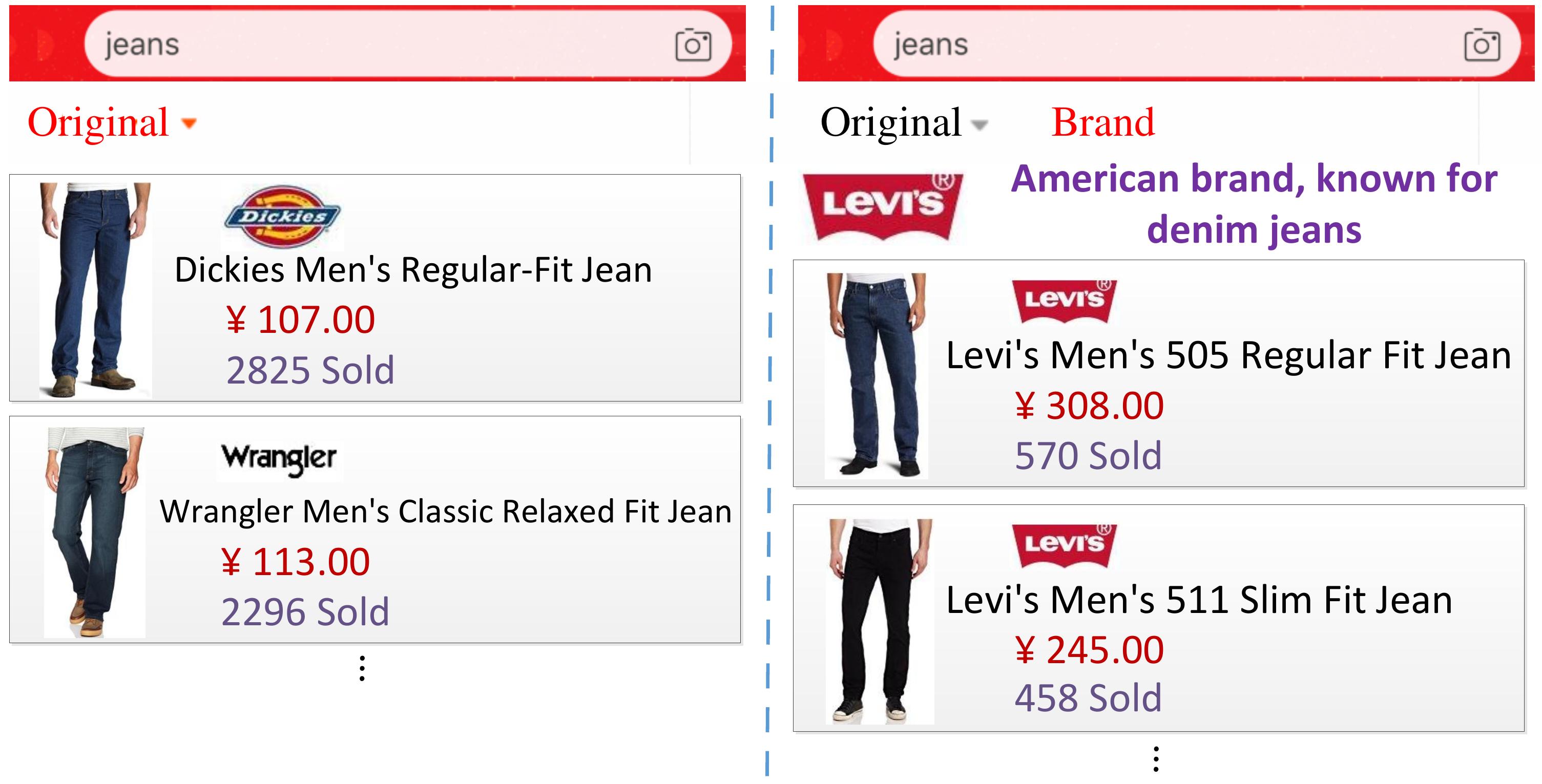}
\end{center}
\vspace*{-15pt}
   \caption{The original ranking system of the platform is shown in the left. Users can switch to our brand-level ranking system by clicking the button ``Brand" as in the right.}
\label{fig:ABtest}
\vspace*{-5pt}
\end{figure}

To better evaluate users' real response to our brand-level ranking system, we conduct a series of online experiments on the Tmall search scenario in Taobao. In this scenario, there are about $5\times 10^8$ \emph{clicks} and $2\times 10^6$ \emph{purchases} on items of nearly $10^5$ brands from over $10^7$
customers within one normal day. A standard A/B test is conducted online. As shown in Figure \ref{fig:ABtest}, we design two user interfaces, one adopts the original ranking system (left), and a button ``Brand" is added to the other one (right), where users can switch to our brand-level ranking system by clicking the button. We preserve the original ranking system in the right for users who have no brand preference. For each setting, the same number (about $5\times 10^5$ per day) of users are randomly selected for A/B test. We perform the online experiments for seven days, and the average CTR, ATIP and GMV (described in section 4.1) per day are reported.

\begin{table}[tb]
\scriptsize
\centering
\begin{tabular}{|c|c|c|c|}
\hline
 &CTR&ATIP&GMV\\ \hline
 Baseline&50.17\%&CNY 140.13&CNY 4824001.84\\ \hline
 New Version&50.37\%&CNY 144.90&CNY 4993468.95\\ \hline
 Improvement&0.39\%&3.40\%&3.51\%\\ \hline
\end{tabular}\vspace*{-5pt}\caption{Performance of Online Experiments (\emph{Baseline} corresponds to the left system of Figure \ref{fig:ABtest} and \emph{New Version} represents the right system. \emph{Improvement} is a relative growth of \emph{New Version} compared to \emph{Baseline}, e.g. $3.40\%\approx (144.90-140.13)/140.13$).}
\vspace*{-10pt}
\label{table:Performance of Online Experiment}
\end{table}

The results are shown in Table \ref{table:Performance of Online Experiment}.
We can see that compared with the baseline, CTR and ATIP in the proposed new ranking system are both improved, which indicates that  by incorporating the brand-level ranking system, users are more likely to click the items and at the same time, they tend to purchase items with higher price (likely to be of higher quality). As a result, the key metric to optimize in this platform, i.e. GMV, has an improvement of 3.51\%. Considering the traffic of the platform, it would result in a significant boost in revenue. Our brand-level ranking system has already gone production on the platform, currently serving as a feature requiring user's proactive opt-in to be activated. Nearly $4\times 10^5$ users actively use this system per day.

\section{Conclusion}
In this paper, we propose a brand-level ranking system to better satisfy the demand of users who have preference bias to certain brands. The core of this system is to solve the personalized brand ranking problem. In our solution, we firstly carefully design the brand features, then rank the brands
by an adapted Attention-GRU model. In future work, we will explore the effectiveness of our modifications on other machine learning models (e.g. Modification 1 on Convolutional Neural Network \cite{krizhevsky:2012imagenet}, Modification 2 on Matrix Factorization \cite{qian:2016non}). In addition, we would improve the design of our ranking model (e.g. considering the exploration of new brands by reinforcement learning).

\section*{Acknowledgements} This work was supported in part by the National Nature Science Foundation of China (Grant Nos: 61751307, 61522206, 61373118, 61672409), the grant ZJU Research 083650 of the ZJUI Research Program from Zhejiang University, the National Youth Top-notch Talent Support Program, the Major Basic Research Project of Shaanxi Province (Grant No. 2017ZDJC-31), and the Science and Technology Plan Program in Shaanxi Province of China (Grant No. 2017KJXX-80). Also thanks to the colleagues in Alibaba Group.

\bibliographystyle{named}
\bibliography{ijcai18}

\end{document}